\documentclass[
aps,%
12pt,%
final,%
notitlepage,%
oneside,%
onecolumn,%
nobibnotes,%
nofootinbib,% 
superscriptaddress,%
noshowpacs,%
centertags]%
{revtex4}
\begin{document}
\selectlanguage{russian}

\title{Моделирование эксклюзивных
партонных распределений\\
 и дальних быстротных
 корреляций в \textit{pp}-столкновениях \\
при энергиях БАК}% Разбиение на строки осуществляется командой \\
%УДК 539.125.17
\author{\firstname{В.~Н.}~\surname{Коваленко}}
% Здесь разбиение на строки осуществляется автоматически или командой \\
\email{nvkinf@rambler.ru}
\affiliation{%
Санкт-Петербургский государственный университет, Россия
}%

%\date{30 ЯНВАРЯ 2012 Г.}
%\today печатает cегодняшнее число`		

\begin{abstract}
Мягкая часть \textit{pp}-взаимодействия рассматривается в рамках феноменологической модели с образованием цветных струн. В предположении, что элементарное столкновение реализуется как взаимодействие двух цветовых диполей, для фиксации параметров модели оценивается полное неупругое сечение и множественность заряженных частиц. Особое внимание уделяется моделированию необходимых для описания корреляций эксклюзивных партонных распределений с учетом закона сохранения энергии и фиксации центра масс. Разработан алгоритм слияния струн в поперечной плоскости, учитывающий их конечную протяженность по быстроте. В рамках этого механизма найдено влияние эффектов слияния струн на дальние
корреляции.
\end{abstract}

\maketitle
\section{Введение}
В настоящее время проводятся эксперименты по изучению столкновений адронов и тяжелых ионов на коллайдере LHC (ЦЕРН) при
сверхвысоких энергиях.
 Подавляющее большинство частиц, рождающихся в таких столкновениях, принадлежат к мягкой составляющей спектра,
 не описываемой в рамках теории возмущений квантовой хромодинамики (КХД), и поэтому используются
  полуфеноменологические модели.

Одной из таких моделей, применяемой для описания мягкой составляющей адронных и ядерных взаимодействий при высоких энергиях, является модель кварк-глюонных струн \cite{Capella,Werner1}, ведущая свое происхождение от реджевского подхода. В этой модели при взаимодействии адронов на первом этапе формируются протяженные объекты -- кварк-глюонные струны. Образование наблюдаемых адронов происходит на втором этапе в процессе фрагментации этих струн. Поскольку кварк-глюонная струна является протяженным объектом, дающим при фрагментации вклад в широкий интервал быстрот, для изучения свойств цветных струн (в том числе возможности их слияния) было предложено использовать так называемые дальние корреляции, соответствующие выбору наблюдаемых из различных быстротных окон, разделенных достаточно большим интервалом быстрот. Эти дальние корреляции в настоящее время являются одним из главных объектов исследования коллектива ALICE/СПбГУ, при этом ведутся как экспериментальные, так и теоретические исследования \cite{lrcA}.

Теоретическое изучение дальних корреляций ведется с помощью метода Монте-Карло \cite{V-C_1,V-C_2},
разработаны алгоритмы \cite{Lakomov1}, описывающие протон-протонные столкновения в рамках
концепции бесконечных по быстроте струн. Однако, для более корректного описания таких характеристик, как корреляции, существенно, в частности, моделировать эксклюзивные партонные распределения, также как и особенности образования цветных струн, в том числе слияния в поперечной плоскости \cite{braun9,braunp}. В связи с этим целью настоящей работы является
моделирование эксклюзивных партонных распределений, а также
 исследование протон-протонных столкновений в рамках модели, учитывающей конечную длину по быстроте
и эффекты слияния кварк-глюнных струн.

\section{Эксклюзивные партонные распределения}
При описании рождения частиц в рамках партонно-струнных моделей обычно используются \cite{7pdf,6pdf}
инклюзивные партонные распределения по доле продольного импульса $x$. Однако непосредственное
применение данного подхода встречается с трудностями при изучении
корреляционных явлений и, в частности, построении монте-карловских моделей,
поскольку в данном случае необходимо иметь информацию о том,
как партоны распределены совместно, т. е. использовать эксклюзивное распределение.
Для построения эксклюзивного партонного распределения по долям импульса мы исходим из того, что
независимость партонов ограничена только выполнением закона сохранения энергии-импульса, следовательно, оно имеет факторизованный вид \cite{bopps}:
\begin{equation}\label{vidrho}
	\rho(x_1, x_2, ... ,x_N)= \rho_1(x_1)\rho_2(x_2)\cdot ... \cdot \rho_N(x_N)\cdot\delta( \sum\limits_{j=1}^N x_j - 1).
\end{equation}
Здесь $x_1,... x_N$ -- доли продольного импульса партонов, при этом
 $N=2n$, где $n$ -- число пар (морских кварк-антикварк либо валентная пара кварк-дикварк), входящих в состав протона.
Множители $\rho_j(x_j)$  определяются из соответствия инклюзивнм структурным
 функциям, имеющим вид \cite{7pdf,6pdf}:
\begin{eqnarray}\label{structfFunctions}
 f_u(x)&=&f_{\bar u}(x)=C_{u,n}\ x^{-\frac{1}{2}}(1-x)^{\frac{1}{2}+n},\\
 f_d(x)&=&f_{\bar d}(x)=C_{d,n}\ x^{-\frac{1}{2}}(1-x)^{\frac{3}{2}+n},\\	   
 f_{ud}(x)&=&C_{ud,n}\ x^{\frac{3}{2}}(1-x)^{-\frac{3}{2}+n},\\
 f_{uu}(x)&=&C_{uu,n}\ x^{\frac{5}{2}}(1-x)^{-\frac{3}{2}+n}.
 \label{structfFunctions1}
\end{eqnarray}	
При этом предполагается, что при $n\geq 2$ морские кварки и антикварки имеют такое же инклюзивное распределение, что и
валентные кварки. В $2/3$ событий дикварк имеет конфигурацию \textit{ud}, в $1/3$ -- \textit{uu}.

Соответствующее эксклюзивное распределение для произвольного числа кварк-антикварковых пар имеет вид
	\begin{equation}\label{vyr}
	\rho(x_1,... x_N)=c\cdot\prod\limits_{j=1}^{N-1} x_j^{-\frac{1}{2}} 
	\cdot x_N^{\alpha_N} \cdot \delta(\sum\limits_{i=1}^N x_i -1),
	\end{equation}
где валентный кварк имеет номер $N-1$, дикварк -- $N$, а остальные номера принадлежат морским кваркам и антикваркам.
$\alpha_N=3/2$ (\textit{ud}-дикварк), $\alpha_N=5/2$ (\textit{uu}-дикварк). Формулы (\ref{structfFunctions}) -- (\ref{structfFunctions1}) можно получить из (\ref{vyr})
путем интегрирования по части переменных.

Для генерации случайных величин (\ref{vyr}) с учетом дельта-функции
необходимо применять особые методы. 
Одним из наиболее общеупотребимых и универсальных является метод Неймана \cite{Bopp1982}, 
однако он неэффективен при большом числе переменных. Явный вид распределения позволяет
значительно повысить эффективность генерации.
Сделаем замену $x_j=y_j^2, j=1, ... \ N-1$. Тогда выражение (\ref{vyr}) для плотности распределения
примет более простой вид:
\begin{equation}\label{vyrost}\nonumber
	\rho(y_1,...,y_{N-1}, x_N)=c\cdot 2^{N-1} \cdot {x_N}^{\alpha_N} \cdot 
 \delta(\sum\limits_{i=1}^{N-1} y_i^2-1+x_N).
%= c_1 \cdot (1-\sum\limits_{i=1}^{N-1} y_i^2)^{\alpha_N}=c_1 (x_N)^{\alpha_N}.
\end{equation}

Произведем еще одну замену: $ y_i=z_i \sqrt{1-x_N}$, $i=1, ... \ N-1$. Тогда

\begin{equation}\label{vzrost}
	\rho(z_1,...,z_{N-1}, x_N)=c\cdot 2^{N-1} \cdot {x_N}^{\alpha_N} (1-x_N)^{(N-3)/2} \cdot 
 \delta(\sum\limits_{i=1}^{N-1} z_i^2-1).
%= c_1 \cdot (1-\sum\limits_{i=1}^{N-1} y_i^2)^{\alpha_N}=c_1 (x_N)^{\alpha_N}.
\end{equation}

Данная формула означает, что $x_N$ подчиняется бета-распределению, а $z_j$ распределены равномерно
на единичной сфере. Пусть $\xi_1... \xi_{N-1}$ -- стандартные нормально распределенные
случайные величины, а $\xi_N$ -- равномерно распределена на отрезке $(0,1)$.
 $x_N=I^{^{-1}}_{\xi_N}(\alpha_N+1,\frac{N-3}{2}+1)$, $x_i=\frac{\xi_i^2}{\sum\limits_{j=1}^{N-1} \xi_j^2}\cdot(1-x_N)$ будут удовлетворять исходному распределению. Здесь $I^{^{-1}}_{x}(a,b)$ --
 обратная регуляризованная неполная бета-функция.
Описанный метод позволяет осуществлять генерацию долей импульсов кварков
	за время, линейно растущее при увеличении $N$.

Для партонной плотности в плоскости прицельного параметра мы используем гауссово распределение \cite{Lakomov1}:
	\begin{equation} \label{raspxy}
\rho(x,y)=\frac{1}{\pi r_0^2} e^{-\frac{\textbf{r}^2}{r_0^2}},
	\end{equation}
Здесь $\textbf{r}=(x,y)$ -- двумерный вектор поперечных координат партона. Параметр $r_0$ связан со среднеквадратичным радиусом протона: $<r_N^2>=\frac{3}{2} r_0^2$.

При генерации положений партонов в плоскости прицельного параметра нельзя считать положения всех
 партонов независимыми, поскольку
необходимо учесть, что в каждом событии центр масс протона должен совпадать с его центром,
определяемым формулой (\ref{raspxy}):
	\begin{equation} \label{pzscm}
\sum\limits_{j=1}^N \textbf{r}_j \cdot E_j =0.
	\end{equation}

Данное условие можно связать с требованием сохранения поперечной
части вектора лоренцева момента, то есть соблюдения
релятивисткого закона движения центра масс во время эволюции протона, предшествующей
столкновению с другим протоном. Действительно, с учетом того, что 
суммарный поперечный импульс партонов равен нулю, поперечная часть вектора
лоренцева момента  $\sum\limits_{j=1}^N (t \textbf{p}_j - \textbf{r}_j E_j) =- \sum\limits_{j=1}^N \textbf{r}_j \cdot E_j =0.$
%Данное требование соответствует сохранению суммарного момента импульса в течении эволюции
%протона, предшествующей столкновению. 

При ультрарелятивистских энергиях $E_j \simeq |p_j| \simeq {p_{\text{z}}}_j  $  (\ref{pzscm})
эквивалентно
	\begin{equation} \label{zscm}
\sum\limits_{j=1}^N  \textbf{r}_j \cdot x_j =0.
	\end{equation}
	
Непосредственным следствием данного условия является то, что в среднем
более тяжелый дикварк находится ближе к центру, а морские кварк-антикварковые
 пары образуются на периферии. Стоит отметить, что к аналогичному выводу пришли
 авторы работы \cite{distmult}, анализируя экспериментальные данные.
	
В итоге эксклюзивное распределение  координат партонов в плоскости прицельного параметра 
должно удовлетворять следующим требованиям:
\begin{enumerate}
	\item центр масс партонов покоится, $\sum\limits_{j=1}^N \textbf{r}_j \cdot x_j =0$ выполнено в каждом событии;
	\item инклюзивное распределение каждого партона является гауссовым;
	\item нормировка: $<r^2>=<\frac{1}{N}\sum\limits_{j=1}^N {r_j}^2>={r_0}^2.$
\end{enumerate}

Генерация координат партонов в плоскости прицельного параметра происходит следующим образом:
\begin{enumerate}
	\item формируется $2N$ независимых
стандартных гауссовых величин $\xi_1, ... \ \xi_N, \eta_1, ... \ \eta_N;$
	\item вычисляются координаты центра масс: ${\text{x}}_{\text{ц.м.}}=\sum\limits_{j=1}^N x_j \xi_j, \ 	{\text{y}}_{\text{ц.м.}}=\sum\limits_{j=1}^N x_j \eta_j;$
	\item производится сдвиг: $\tilde{\xi_i}=\xi_i-{\text{x}}_{\text{ц.м.}}, \ \tilde{\eta_i}=\eta_i-{\text{y}}_{\text{ц.м.}};$
	\item масштабирование с постоянным коэффициентом: ${\text{x}}_i=\tilde{\xi_i}/a, \ {\text{y}}_i=\tilde{\eta_i}/a.$
\end{enumerate}
	Полученные величины $({\text{x}}_i,$ ${\text{y}}_i)$ являются координатами партонов.
	
 Остановимся подробнее на вычислении величины $a$ (константы, зависящей только от 
 количества партонов и от типа дикварка) из условия нормировки.
	Координаты для кварка и для дикварка до масштабирования задаются следующим образом:
\begin{eqnarray}\label{vqq}
	\tilde\xi_{j}&=&(1-x_{j})\xi_{j}-x_{N}\xi_{N}-\sum\limits_{i=1, i \neq j}^{N-1}x_i\xi_{i} \ ,\  j=1, 2, ...\ N-1, \\
	\tilde\xi_{N}&=&(1-x_{N})\xi_{N}-\sum\limits_{i=1}^{N-1}x_i\xi_{i}.	 
\end{eqnarray}	
	Подсчитаем дисперсии, учитывая независимость величин $\xi_j$  и $x_i$:
\begin{eqnarray}\label{disps}
	 D\tilde\xi_q &=&<(1-x_q)^2>+<x_{qq}^2>+(N-2)<x_q^2>,\\
	 D\tilde\xi_{qq}&=&<(1-x_{qq})^2>+(N-1)<x_q^2>.
	 \label{disps1}
\end{eqnarray}	

	Для нахождения величины $a$
	подставим ${\text{x}}_i=\tilde{\xi_i}/a, \ {\text{y}}_i=\tilde{\eta_i}/a$ в
	условие нормировки, а также учтем, что в протоне один дикварк и $N-1$ кварков (и антикварков).
	Тогда уравнение для нахождения $a$ примет вид
\begin{equation}\label{finda}
	N\cdot a^2 =  D\tilde{\xi_{qq}} + (N-1) D\tilde{\xi_q} .
\end{equation}	

Явные выражения для (\ref{disps}), (\ref{disps1})
получаются путем усреднения c помощью инклюзивных распределений
 (\ref{structfFunctions}) -- (\ref{structfFunctions1}), а последующая
 подстановка в (\ref{finda}) дает окончательное выражение для $a$:

\begin{equation}
\label{alphaud} a_{ud}^2={\frac{(2n-1)(n^2+6n+12)}{2n(n+2)(n+3)}},
\hspace*{1cm} a_{uu}^2={\frac{(2n-1)(n^2+8n+24)}{2n(n+3)(n+4)}}.
\end{equation}	
Здесь $a_{ud}$ обозначена величина $a$ для конфигурации дикварка $ud$, $a_{uu}$ -- для конфигурации дикварка $uu$.

%Заметим, что соблюдение условий нормировки контролировалось в ходе монте-карловских симуляций.

%	в симуляциях контролировать выполнение $<\frac{1}{N}\sum\limits_{j=1}^N {r_j}^2>={r_0}^2$.

%Очень кратко про гаусс, про радиус, про сохранение момента, описание генерации (кратко).

\section{Описание модели}

\subsection{Цветные диполи}
Для формулировки модели неупругих протон-протонных столкновений мы предполагаем,
 что элементарное столкновение партонов реализуется как взаимодействие двух цветовых диполей,
состоящих из валентного кварка и дикварка, либо из кварк-антикварковой пары.
Амплитуда
вероятности взаимодействия двух диполей с координатами $( \textbf{r}_1 \textbf{r}_2)$ и $(\textbf{r}_3  \textbf{r}_4)$ 
в плоскости прицельного параметра имеет следующий вид \cite{Gustafson}: 
\begin{equation} \label{withlog}
	f=\frac{\alpha_S^2}{8}\ln^2 \frac {( \textbf{r}_1 - \textbf{r}_3 )^2 ( \textbf{r}_2 - \textbf{r}_4 )^2 }
									{( \textbf{r}_1 - \textbf{r}_4 )^2 ( \textbf{r}_2 - \textbf{r}_3 )^2 },
\end{equation}
%где $a_S$ -- константа.
В рамках приближения эйконала вероятность взаимодействия двух диполей 
\begin{equation}\label{pij}
{p_{ij}=1-e^{-f_{ij}}}.
\end{equation}
Полная вероятность неупругого взаимодействия двух протонов ${p=1-e^{-\sum\limits_{i,j} f_{ij}}}$,
где суммирование ведется по всем диполям.
Величина $\alpha_S$ имеет смысл эффективной константы взаимодействия,
и ее значение используется как подгоночный параметр для более правильного 
описания экспериментальных данных, причем предполагается, что она не зависит ни от энергии, 
ни от количества кварк-антикварковых пар в протоне.

Влияние конфайнмента может быть учтено следующим образом \cite{Lonnbland1,Gustafson}:
заменить кулоновский пропагатор $\Delta( \textbf{r})=\int \frac{d^2 \textbf{k}}{(2\pi)^2} \frac{e^{i  \textbf{k}\cdot  \textbf{r}}}{k^2}$, который получается из (\ref{withlog}), на соответствующий пропагатор Юкавы:  $\frac{1}{k^2+M^2}$, где $M=1/r_{\text{max}}$ -- характерный масштаб конфайнмента. Тогда формула (\ref{withlog}) примет вид:
		\begin{equation} \label{newformula}
		\nonumber
			f=\frac{\alpha_S^2}{2}\Big[ K_0\left(\frac{| \textbf{r}_1- \textbf{r}_3|}{r_{\text{max}}}\right) +
			K_0\left(\frac{| \textbf{r}_2- \textbf{r}_4|}{r_{\text{max}}}\right) 
			- K_0\left(\frac{| \textbf{r}_1- \textbf{r}_4|}{r_{\text{max}}}\right)
			- K_0\left(\frac{| \textbf{r}_2- \textbf{r}_3|}{r_{\text{max}}}\right)	\Big]^2.
		\end{equation}		
Согласно оценкам \cite{Braun-Vechernin}, отношение квадратов радиусов протона и адрона должны быть порядка $\frac{1}{10}$. В этом случае $r_{\text{max}}\simeq 0.2 - 0.3$ фм.
Поскольку при $r \ll r_{\text{max}}$ $K_0(r/r_{\text{max}})\simeq-\ln(r/(2r_{\text{max}}))$, то в этом предельном случае
мы снова возвращаемся к исходной формуле (\ref{withlog}). При $r \gg r_{\text{max}}$ модифицированная
функция Бесселя ведет себя следующим образом:
\begin{equation} \label{modBesself}
	K_0\left(\frac{r}{r_{\text{max}}}\right)\simeq \sqrt{\frac{\pi r_{\text{max}}}{2r}}e^{-\frac{r}{r_{\text{max}}}},
\end{equation}	 
что  обеспечивает экспоненциальное падение вероятности столкновения двух протонов при большом прицельном параметре.

\subsection{Вычисление наблюдаемых величин}
Вероятности столкновений диполей, полученные ранее, используются для построения матрицы столкновений.
Каждый диполь может взаимодействовать только с одним другим диполем, поэтому заполнение матрицы столкновений
начинается с валентных диполей.

Следующим этапом является генерация струн в пространстве быстроты. 
Величины $y_{\text{min}}, y_{\text{max}}$ (концы струны) определяются из кинематического условия
распада струны только на две частицы со средним $p_t=0.3$ ГэВ
и массами $m_{\pi}=0.15$ ГэВ (для пиона, если на конце струны кварк или антикварк) или $m_{p}=0.94$ ГэВ (для нуклона, если на конце струны дикварк).
Слишком короткие струны исключаются из рассмотрения требованием, чтобы
сумма масс рожденных частиц была меньше, чем масса струны, равная $\sqrt{s \cdot x_A \cdot x_B}$, где $x_A, x_B$ -- доли импульса партонов на концах струны. Поперечные координаты центра струны полагаются равными среднему арифметическому 
соответствующих координат партонов на ее концах.

Для вычисления распределения множественности и среднего поперечного импульса по быстроте предполагается,
что множественность на единицу быстроты ($\mu_0$) и средний поперечный импульс ($p_0$) от одной струны фиксированы и не зависят
ни от энергии, ни от других параметров модели. При большем количестве струн происходит их перекрытие
в плоскости прицельного параметра, что приводит к необходимости учета
взаимодействия между ними. Данное взаимодействие может быть выполнено в модели, 
принимающей во внимание эффекты слияния струн \cite{braun9}.

Для нахождения множественности и среднего поперечного импульса,
с учетом эффекта слияния цветных струн был использован вариант с локальным %translate: string fusion%
слиянием (overlaps) в рамках дискретной модели, предложенной в \cite{diskr1,diskr2,diskr3}.
В этом случае согласно \cite{braun9} предполагается, что средняя множественность на единицу быстроты и средний поперечный импульс заряженных частиц, излученных из области перекрытия $k$ струн, описываются
следующими выражениями:
\begin{equation} \label{muptloc}
	\left\langle \mu\right\rangle_k=\mu_0 \sqrt{k} \frac{S_k}{\sigma_0}, \hspace*{1cm}
	\left\langle p_t^2\right\rangle_k=p_0 \sqrt{k}. 
\end{equation}	 
Здесь $S_k$ -- поперечная площадь области, где произошло перекрытие $k$ цветных струн,
 $\sigma_0$ -- поперечная площадь струны, $\mu_0$ и $p_0$ -- средняя множественность на единицу
быстроты и средний поперечный импульс заряженных частиц, когда они рождаются
от распада одиночной струны.
В дискретной модели поперечная плоскость заменяется решеткой с площадью ячейки, равной площади струны, и струны считаются слившимися, если их центры попадают в одну и ту же ячейку.
Для учета конечной длины струн каждая ячейка делится на интервалы по
быстроте так, чтобы количество струн в каждом интервале было постоянным. После этого
расчеты производятся с дополнительным суммированием по всем быстротным интервалам.

Для вычисления $nn$ и $p_tn$ корреляционных функций 
для дальних (с разнесенными по быстроте окнами $F$ и $B$)
используется метод \cite{V-C_1,Lakomov1}, в
котором для каждой струнной конфигурации, т.е. для каждого события, сгенерированного методом Монте-Карло, вычисляется
средняя множественность и средний поперечный импульс с последующим усреднением по всем конфигурациям:
\begin{eqnarray}
\langle {n}_B \rangle_{n_F} = \frac{ \sum\limits_C { w(C) \  \langle {n}_B \rangle_{C} \ P_C(n_F)} }
  { \sum\limits_C { w(C) P_C(n_F)}}, \\
\langle {p_t^2}_B \rangle_{n_F} = \frac{ \sum\limits_C { w(C) \ \langle {p_t^2}_B \rangle_{C} \ P_C(n_F)  } }
  { \sum\limits_C { w(C) P_C(n_F)}}, 
\end{eqnarray}
где $w(C)$ -- вероятность реализации конфигурации $C$. В качестве вероятности $P_C(n_F)$ иметь множественность ${n_F}$ 
в переднем окне при данной конфигурации, как и в работах  \cite{V-C_1,Lakomov1}, используется
распределение Пуассона со средним значением, равным $\langle {n}_F \rangle_{C}$.

Полное неупругое сечение в настоящей работе вычисляется двумя способами:
\begin{eqnarray} 
\label{osigma}	\sigma^{(1)}&=&\frac{\sum\limits_{\text{sim}} p}{N_{\text{sim}}} \cdot a^2 ,
\\
\label{dsigma}	\sigma^{(2)}&=&\frac{N_{\text{ev}}}{N_{\text{sim}}} \cdot a^2 ,
\end{eqnarray}			 		
где $a^2$ -- площадь области в плоскости прицельного параметра, в которой генерируются протоны,
$p$ -- полная вероятность неупругого столкновения двух протонов в каждом событии,
$N_{\text{sim}}$ -- общее число симуляций, $N_{\text{ev}}$ -- число симуляций,
в которых было хотя бы одно столкновение с образованием струн; суммирование ведется
по всем симуляциям.
Второй способ представляет собой обычный метод в монте-карловских симуляциях, 
в то время как первый является следствием модели цветовых диполей.
Поскольку обе формулы дают один и тот же результат, первый способ является основным, второй используется для контроля.

\section{Монте-Карловский алгоритм}
Моделирование процесса \textit{pp}-столкновения начинается с генерации координат
 $ \textit{r}_A,  \textit{r}_B $ центров протонов в плоскости прицельного параметра
с равномерным распределением в квадрате площадью $a^2$ (minimum bias симуляции),
причем величина $a$ выбирается много больше, чем характерный радиус взаимодействия протонов, 
так, чтобы при ее дальнейшем увеличении результаты от нее не зависели.

Далее генерируются число диполей для каждого протона ($m_A$, $m_B$)
согласно распределению Пуассона с параметром $\lambda$; случай $m=0$ отбрасывается,
поскольку диполь из валентных дикварка и кварка обязательно должен присутствовать.
Количество партонов ($N$) принимается равным удвоенному числу диполей.
Дикварку приписывается тип $uu$ с вероятностью $1/3$ или $ud$ -- с вероятностью $2/3$.

Затем генерируются доли импульса партонов (\ref{vyr}) и поперечные координаты.
Параметр $r_0$ выбирается равным $0.51$ фм, исходя из среднеквадратичного радиуса протона $r_N = 0.63$ фм.
Из пар кварк-антикварк и кварк-дикварк строится набор диполей для каждого протона и
генерируется матрица взаимодействия с помощью вероятностей (\ref{pij}, \ref{newformula}).
Вычисляется полная вероятность взаимодействия $p=1-e^{-\sum\limits_{ij} f_{ij}}$,
 которая используется для подсчета сечения.

По матрице столкновений строится набор струн, и с использованием дискретного алгоритма слияния производится
вычисление распределений множественности и среднего поперечного импульса по быстроте, по которым
рассчитываются величины дальних корреляций, а также средняя множественность и средний поперечный импульс для заданных
быстротных окон.

	\section{Процедура фиксации параметров}	\label{proc}	

	Рост множественности и сечения в данной модели обеспечивается
	за счет увеличения количества кварк-антикварковых пар (параметра $\lambda$).
Величина $ \lambda $ фиксируется из экспериментальных данных 
по полному неупругому сечению \cite{sigmafit}:
	\begin{eqnarray} \label{sigmafit1}
		  \sigma^{in}_{pp} (s) = 32.08 - 1.574 \cdot \ln (s) + 0.6622 \cdot \ln^2(s).
	\end{eqnarray}	
В связи с тем, что расчет сечения в рамках данной модели при заданном $\lambda$ 
не использует величину энергии, симуляции проводятся в два этапа: на первом
устанавливается связь $\sigma^{in}_{pp}$ и $\lambda$, а на втором для заданной энергии 
определяется $\lambda$ c использованием фита (\ref{sigmafit1}) и проводятся остальные расчеты.

Величины $\mu_0$ и $p_0$ фиксируются в одной точке при низкой энергии.
	
Экспериментальные данные для множественности
и поперечного импульса были взяты из работ \cite{alice10,CMSfit}:
\begin{equation}
 \langle p_t \rangle = 0.413 - 0.0171 \ln{s} + 0.00143 \ln^2{s}, \\
\end{equation}
\begin{equation}
 \frac{dN}{d\eta}|_{\eta=0} = 0.815 \cdot s^{0.10671}.
\end{equation}
	
	Другие параметры модели выбирались следующим образом:
    для $r_{\text{max}}\simeq 0.2 - 0.3$ фм отношение $r_{\text{max}} / r_0$ выбиралось равным $0.5$;
    константа $\alpha_S$ выбиралась для наилучшего описания зависимости множественности от энергии.
    Радиус струны $r_{\text{str}}=0.3$ фм соответствует оценкам \cite{braunp,V-C_1,V-C_2,diskr1,diskr2,diskr3},
    полученным при описании поперечного импульса и дальних корреляций в ядро-ядерных взаимодействиях.

\section{Полученные результаты} 
\begin{figure}[t!]
\setcaptionmargin{5mm}

%\onelinecaptionsfalse % если подпись к рисунку многострочная
\onelinecaptionstrue  % если подпись к рисунку однострочная
{
\includegraphics[width=120mm]{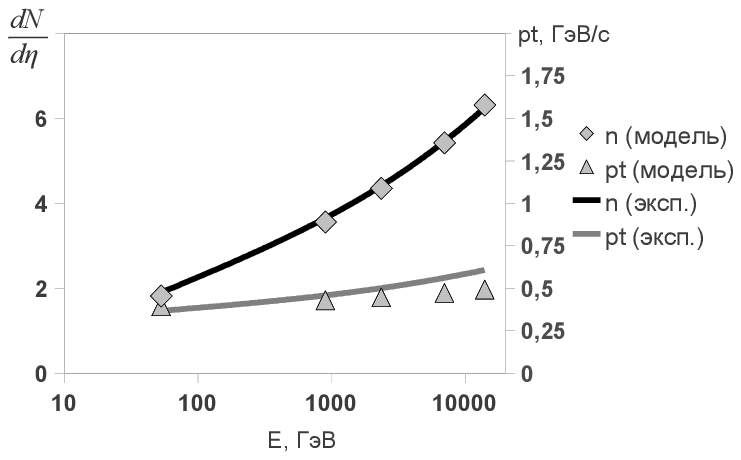}
\captionstyle{normal} \caption{\label{pic1} Множественность и средний поперечный импульс заряженных частиц в \textit{pp}-столкновениях. Линиями показаны фиты экспериментальных данных из работ \cite{alice10,CMSfit}.}
}

\end{figure}

На  рис. \ref{pic1} представлены экспериментальные данные и результаты расчетов среднего
поперечного импульса и множественности заряженных частиц в \textit{pp}-столкновениях
при различных энергиях в модели со слиянием струн ($r_{\text{str}}=0.3$ фм),
при этом использовались следующие параметры:
$\alpha_S=1.4$, $\mu_0=1.02$, $p_0=0.37$ ГэВ. Величина $\lambda$ изменялась от $5.8$ для $\sqrt{s}=50$ ГэВ до $9.6$ для $\sqrt{s}=900$ ГэВ и $16.1$ для $\sqrt{s}=7000$ ГэВ.

Как и в случае без слияния \cite{KovSP}, за счет
подбора параметров ($\alpha_s, \mu_0$)
 удается полностью описать
множественность заряженных частиц в широком диапазоне энергий,
однако при любом их разумном выборе
 рассчитанный рост поперечного импульса
оказывается более медленным, чем наблюдаемый в эксперименте.

\begin{figure}[t!]
\setcaptionmargin{5mm}

%\onelinecaptionsfalse % если подпись к рисунку многострочная
\onelinecaptionstrue  % если подпись к рисунку однострочная
{
\includegraphics[width=100mm]{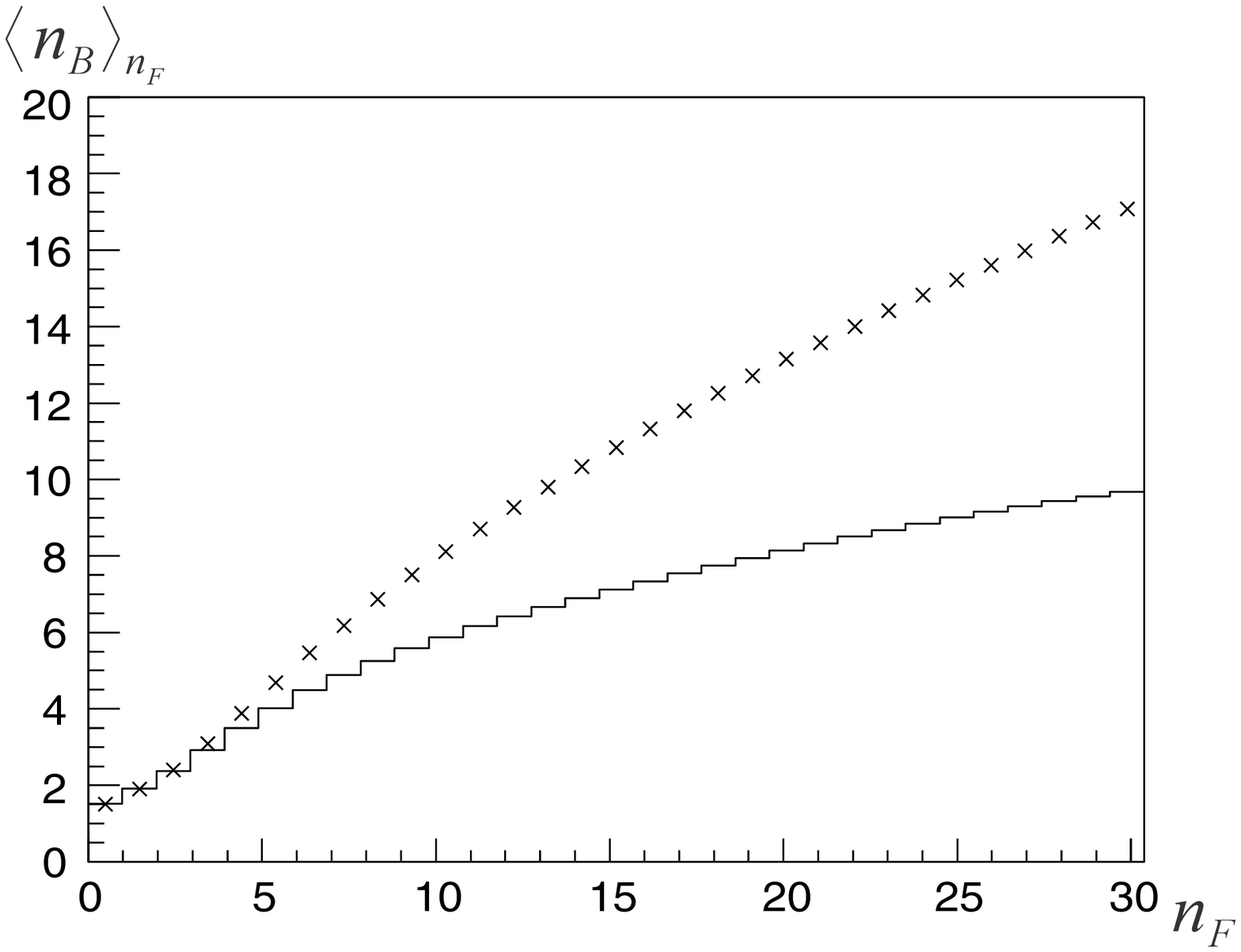}
\captionstyle{normal} \caption{\label{pic2}
$nn$-корреляционная функция в \textit{pp}-столкновениях при энергии $E=900$~ГэВ, вычисленная
в данной работе для быстротных окон $(-0.8;0)$ и $(0.0;0.8)$.
Крестиками показана $nn$-корреляция без учета эффекта слияния струн.
}

}

\end{figure}

На рис. \ref{pic2} представлена $nn$-корреляционная функция, рассчитанная
для энергии ${E=900}$~ГэВ для раздельных окон $(-0.8, 0)$, $(0, 0.8)$.
Нелинейность корреляционной функции имеет место
и в случае отсутствия слияния, однако при включении слияния струн
она становится сильнее. Такое поведение корреляционной функции
находится в согласии с расчетами для ядро-ядерных столкновений \cite{V-C_1,NA49}.

\begin{figure}[t!]
\setcaptionmargin{5mm}

%\onelinecaptionsfalse % если подпись к рисунку многострочная
\onelinecaptionstrue  % если подпись к рисунку однострочная
{
\includegraphics[height=53mm]{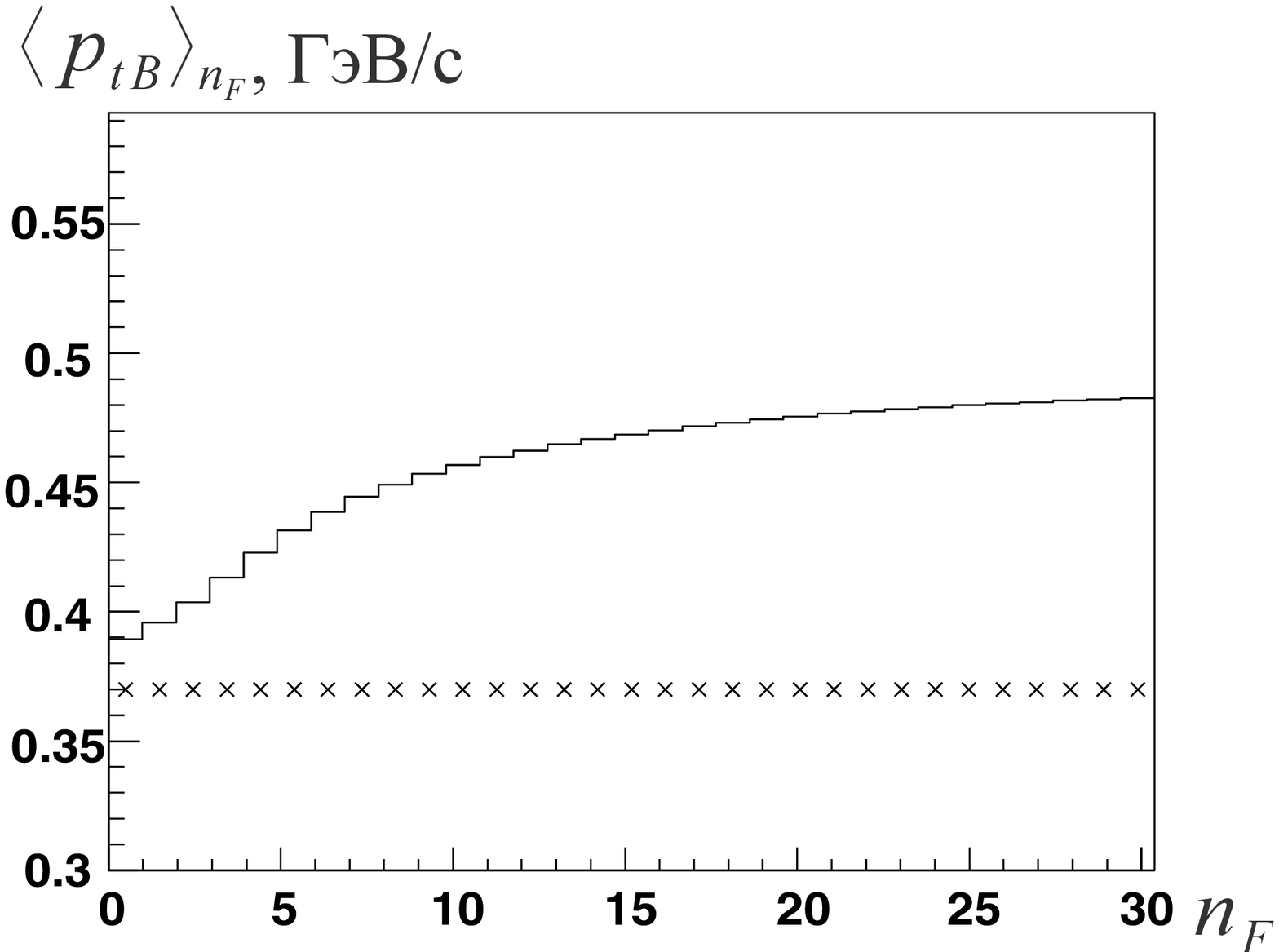}
\includegraphics[height=53mm]{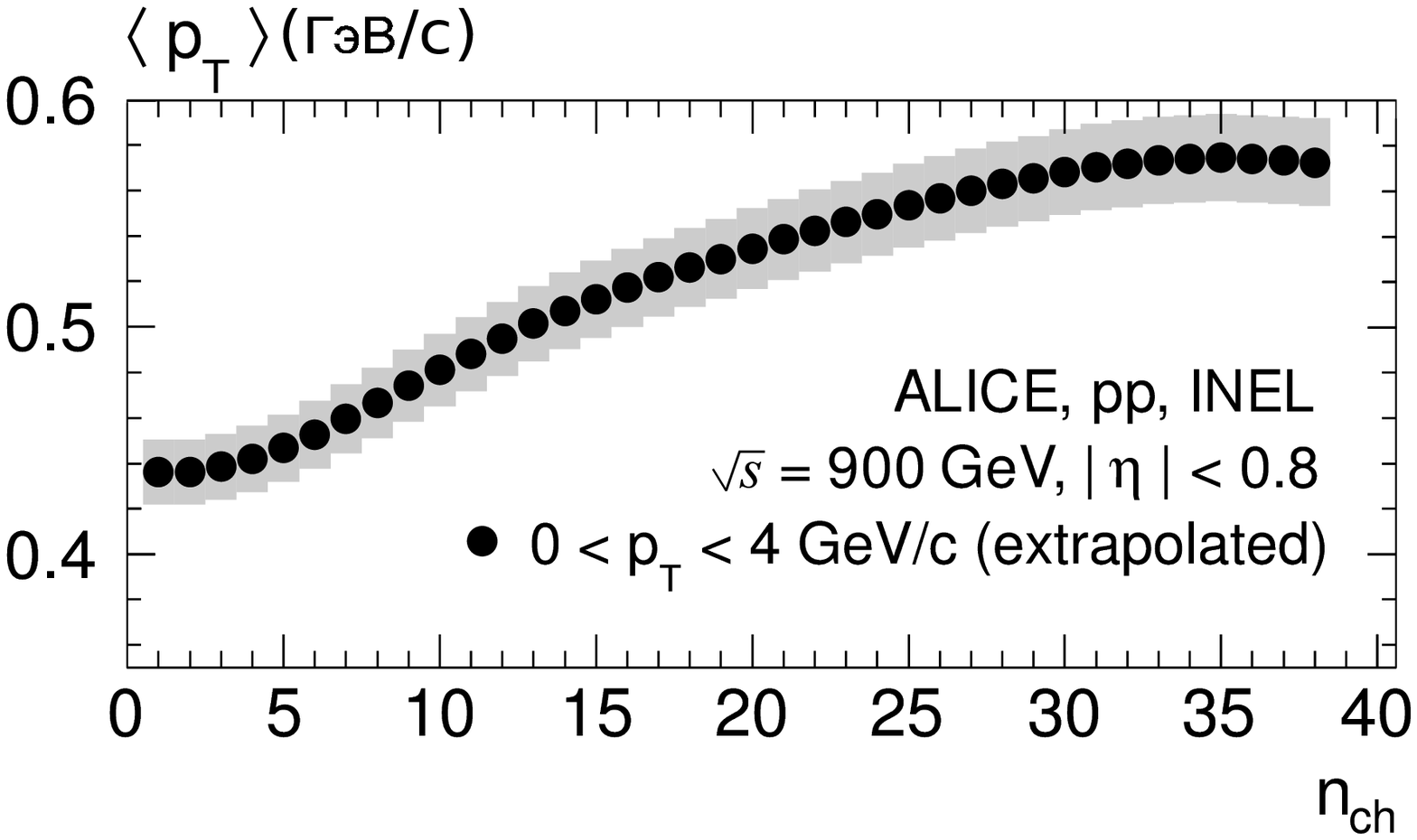}
\captionstyle{normal} \caption{\label{pic3}
Корреляционная функция дальних $p_tn$-корреляций в $pp$-столкновениях, вычисленная в данной работе (слева), для быстротных окон $(-0.8;0)$ и $(0.0;0.8)$;
экспериментальные данные (справа) при энергии $E=900$~ГэВ \cite{Alice900}.
Крестиками показана $p_tn$-корреляция без учета эффекта слияния струн.
}
}

\end{figure}

На рис. \ref{pic3} показаны результаты расчета $p_tn$-корреляционной функции для
тех же быстротных окон.
Для сравнения с экспериментальными данными мы использовали
$\langle {p_t}_B \rangle_{n_F}=\sqrt{\langle {p_t^2}_B \rangle_{n_F}}$.
 К сожалению, пока имеются экспериментальные
данные только о $p_tn$-корреляциях в одном быстротном окне, и сравнение с
представленными данными носит несколько условный характер. В целом
имеется качественное воспроизведение формы корреляционной функции.

Данные по полному неупругому сечению не приводятся, так как эта величина
всегда фиксируется. 

\section{Заключение}

В данной работе предложена модель,  описывающая характеристики
протон-протонных столкновений при высоких энергиях,
которая включает детальное моделирование партонных распределений;
 соответствующие монте-карловские процедуры разработаны для произвольного числа партонов в адроне.
 Это позволило применить алгоритм слияния струн в поперечной плоскости 
 с учетом их конечной протяженности по быстроте для \textit{pp-}рассеяния.
  В рамках этого механизма найдено влияние эффектов слияния струн на дальние по быстроте корреляции между множественностями заряженных частиц, а также между множественностью и средним поперечным импульсом.

В результате показано, что при использовании общепринятого значения радиуса
струны $0.3$ фм рассчитанный рост среднего поперечного импульса с энергией оказывается
меньше экспериментально наблюдаемого. Возможно, это связано
с тем, что на рост поперечного импульса при более высоких энергиях начинают
 оказывать влияние и другие механизмы, такие, как вклад жестких процессов.
Увеличение радиуса струны до $0.4$ фм должно улучшить
количественное описание увеличения поперечного импульса,
однако это может противоречить данным по ядро-ядерным столкновениям \cite{V-C_1, braun9}.

В дальнейшем в рамках предложенной модели
 также планируется изучать зависимость характеристик от 
 ширины и конфигурации быстротных окон.
Алгоритмы, предложенные в данной работе, могут быть применены
к описанию нуклон-нуклонных столкновений при изучении столкновений тяжелых ионов.

Автор благодарен В. В. Вечернину и Г. А. Феофилову
за многочисленные обсуждения и ценные замечания.

\selectlanguage{russian}
%\section*{maik.bst}
%\nocite{*}
%\bibliographystyle{unsrt}
\bibliography{paper.bib}

\begin{thebibliography}{24}
\expandafter\ifx\csname natexlab\endcsname\relax\def\natexlab#1{#1}\fi
\expandafter\ifx\csname bibnamefont\endcsname\relax
  \def\bibnamefont#1{#1}\fi
\expandafter\ifx\csname bibfnamefont\endcsname\relax
  \def\bibfnamefont#1{#1}\fi
\expandafter\ifx\csname citenamefont\endcsname\relax
  \def\citenamefont#1{#1}\fi
\expandafter\ifx\csname url\endcsname\relax
  \def\url#1{\texttt{#1}}\fi
\expandafter\ifx\csname urlprefix\endcsname\relax\def\urlprefix{URL }\fi
\providecommand{\bibinfo}[2]{#2}
\providecommand{\eprint}[2][]{\url{#2}}

\bibitem[{\citenamefont{{A. Capella, U. P. Sukhatme, C. I. Tan, and Van J. Tran
  Thanh}}(1994)}]{Capella}

%\refitem{article}
\bibinfo{author}{\bibnamefont{{A. Capella, U. P. Sukhatme, C. I. Tan, and Van J. Tran
  Thanh}}}, \bibinfo{journal}{Phys. Rept.} \textbf{\bibinfo{volume}{236}},
  \bibinfo{pages}{225} (\bibinfo{year}{1994}).

\bibitem[{\citenamefont{Drescher \emph{et~al.}}(2001)\citenamefont{Drescher,
  Hladik, Ostapchenko, Pierog, and Werner}}]{Werner1}

%\refitem{article}
\bibinfo{author}{\bibfnamefont{H.~J.} \bibnamefont{Drescher}},
  \bibinfo{author}{\bibfnamefont{M.}~\bibnamefont{Hladik}},
  \bibinfo{author}{\bibfnamefont{S.}~\bibnamefont{Ostapchenko}},
  \bibinfo{author}{\bibfnamefont{T.}~\bibnamefont{Pierog}}, \bibnamefont{and}
  \bibinfo{author}{\bibfnamefont{K.}~\bibnamefont{Werner}},
  \bibinfo{journal}{Phys. Rept.} \textbf{\bibinfo{volume}{350}},
  \bibinfo{pages}{93} (\bibinfo{year}{2001}), \eprint{arXiv: hep-ph/0007198}.

\bibitem[{\citenamefont{{B.~Alessandro, et al. (ALICE Collaboration)}}(2006)}]{lrcA}

%\refitem{article}ALICE Collaboration et al 2006 J. Phys. G: Nucl. Part. Phys. 32 1295
\bibinfo{author}{\bibnamefont{{B.~Alessandro, et al. (ALICE Collab.)}}},
  \bibinfo{journal}{ALICE: Physics Performance Report, Vol. II,} 
  \bibinfo{journal}{J.~Phys.~G}
  \textbf{\bibinfo{volume}{32}}, \bibinfo{pages}{1295}
  \bibinfo{year}{(Section 6.5.15: Long-range correlations, p. 1749-1751)}
  (\bibinfo{year}{2006}).

\bibitem[{\citenamefont{Вечернин, Колеватов}(2007{\natexlab{a}})}]{V-C_1}

%\refitem{article}
\bibinfo{author}{\bibfnamefont{В.~В.}~\bibnamefont{Вечернин,}} \bibnamefont{}
  \bibinfo{author}{\bibfnamefont{Р.~С.}~\bibnamefont{Колеватов, ЯФ \textbf{70}, 1846 (2007)}}
  \bibinfo{journal}{[Phys. Atom. Nucl.} \textbf{\bibinfo{volume}{70}},
  \bibinfo{pages}{1797} (\bibinfo{year}{2007}{\natexlab{a}})].
  %\urlprefix\url{http://dx.doi.org/10.1134/S1063778807100158}

%\bibitem[{\citenamefont{N. Armesto, D. Derkach, G. A. Feofilov}(2008{\natexlab{a}})}]{ADF}

%\refitem{article}
%\bibinfo{author}{\bibfnamefont{N.}~\bibnamefont{Armesto,}} \bibnamefont{}
%\bibinfo{author}{\bibfnamefont{D.}~\bibnamefont{Derkach,}} \bibnamefont{}
%  \bibinfo{author}{\bibfnamefont{G. A.}~\bibnamefont{Feofilov, ЯФ 71, 2122 (2007)}}
%  \bibinfo{journal}{[Phys. Atom. Nucl.} \textbf{\bibinfo{volume}{71}},
%  \bibinfo{pages}{2087} (\bibinfo{year}{2008}{\natexlab{a}})].
  %\urlprefix\url{http://dx.doi.org/10.1134/S1063778807100158}


\bibitem[{\citenamefont{Vechernin and Kolevatov}(2007{\natexlab{b}})}]{V-C_2}

%\refitem{article}
\bibinfo{author}{\bibfnamefont{В.~В.}~\bibnamefont{Вечернин,}} \bibnamefont{}
  \bibinfo{author}{\bibfnamefont{Р.~С.}~\bibnamefont{Колеватов, ЯФ \textbf{70}, 1858 (2007)}}
  \bibinfo{journal}{[Phys. Atom. Nucl.} \textbf{\bibinfo{volume}{70}},
  \bibinfo{pages}{1809} (\bibinfo{year}{2007}{\natexlab{a}})].
  %\urlprefix\url{http://dx.doi.org/10.1134/S1063778807100158}


\bibitem[{\citenamefont{{В. В. Вечернин, И. А. Лакомов, А. М.
  Пучков}}(2010)}]{Lakomov1}

%\refitem{article}
\bibinfo{author}{\bibnamefont{{В. В. Вечернин, И. А. Лакомов, А. М. Пучков}}},
  \bibinfo{journal}{Вестн. СПбГУ. Сер. 4. Физика. Химия. Вып. 3, 3}  (\bibinfo{year}{2010}).

\bibitem[{\citenamefont{Braun and Pajares}(2000)}]{braun9}

%\refitem{article}
\bibinfo{author}{\bibfnamefont{M.}~\bibnamefont{Braun}} \bibnamefont{and}
  \bibinfo{author}{\bibfnamefont{C.}~\bibnamefont{Pajares}},
  \bibinfo{journal}{Eur. Phys. J. (C)} \textbf{\bibinfo{volume}{16}},
  \bibinfo{pages}{349} (\bibinfo{year}{2000}).
  %\urlprefix\url{http://dx.doi.org/10.1007/s100520050027}.

\bibitem[{\citenamefont{Armesto \emph{et~al.}}(1996)\citenamefont{Armesto,
  Braun, Ferreiro, and Pajares}}]{braunp}

%\refitem{article}
\bibinfo{author}{\bibfnamefont{N.}~\bibnamefont{Armesto}},
  \bibinfo{author}{\bibfnamefont{M.~A.} \bibnamefont{Braun}},
  \bibinfo{author}{\bibfnamefont{E.~G.} \bibnamefont{Ferreiro}},
  \bibnamefont{and} \bibinfo{author}{\bibfnamefont{C.}~\bibnamefont{Pajares}},
  \bibinfo{journal}{Phys. Rev. Lett.} \textbf{\bibinfo{volume}{77}},
  \bibinfo{pages}{3736} (\bibinfo{year}{1996}),
  %\urlprefix\url{http://link.aps.org/doi/10.1103/PhysRevLett.77.3736}.

\bibitem[{\citenamefont{Aurenche \emph{et~al.}}(1984)\citenamefont{Aurenche,
  Bopp, and Ranft}}]{bopps}

%\refitem{article}
\bibinfo{author}{\bibfnamefont{P.}~\bibnamefont{Aurenche}},
  \bibinfo{author}{\bibfnamefont{F.~W.} \bibnamefont{Bopp}}, \bibnamefont{and}
  \bibinfo{author}{\bibfnamefont{J.}~\bibnamefont{Ranft}},
  \bibinfo{journal}{Z. Phys. C}
  \textbf{\bibinfo{volume}{23}}, \bibinfo{pages}{67} (\bibinfo{year}{1984}).
  %\urlprefix\url{http://dx.doi.org/10.1007/BF01558042}.

\bibitem[{\citenamefont{Kaidalov and Piskunova}(1986)}]{7pdf}

%\refitem{article}
\bibinfo{author}{\bibfnamefont{A.~B.} \bibnamefont{Kaidalov}} \bibnamefont{and}
  \bibinfo{author}{\bibfnamefont{O.~I.} \bibnamefont{Piskunova}},
  \bibinfo{journal}{Z. Phys. C}
  \textbf{\bibinfo{volume}{30}}, \bibinfo{pages}{145} (\bibinfo{year}{1986}),
  %\urlprefix\url{http://dx.doi.org/10.1007/BF01560688}.

\bibitem[{\citenamefont{Arakelyan \emph{et~al.}}(2002)\citenamefont{Arakelyan,
  Capella, Kaidalov, and Shabelski}}]{6pdf}

%\refitem{article}
\bibinfo{author}{\bibfnamefont{G.~H.} \bibnamefont{Arakelyan}},
  \bibinfo{author}{\bibfnamefont{A.}~\bibnamefont{Capella}},
  \bibinfo{author}{\bibfnamefont{A.~B.} \bibnamefont{Kaidalov}},
  \bibnamefont{and} \bibinfo{author}{\bibfnamefont{Y.~M.}
  \bibnamefont{Shabelski}}, \bibinfo{journal}{Eur. Phys. J. (C)}
  \textbf{\bibinfo{volume}{26}}, \bibinfo{pages}{81} (\bibinfo{year}{2002})
  %\urlprefix\url{http://arxiv.org/abs/hep-ph/0103337v2}.

\bibitem[{\citenamefont{Bopp and Aurenche}(1982)}]{Bopp1982}

%\refitem{article}
\bibinfo{author}{\bibfnamefont{F.~W.} \bibnamefont{Bopp}} \bibnamefont{and}
  \bibinfo{author}{\bibfnamefont{P.}~\bibnamefont{Aurenche}},
  \bibinfo{journal}{Z. Phys. C}
  \textbf{\bibinfo{volume}{13}}, \bibinfo{pages}{205} (\bibinfo{year}{1982}).
  %\urlprefix\url{http://dx.doi.org/10.1007/BF01575773}.

\bibitem[{\citenamefont{distmult}(2004)}]{distmult}

%\refitem{article}
\bibinfo{author}{\bibfnamefont{W. D.}~\bibnamefont{Walker}},
  \bibinfo{journal}{Phys. Rev. D} \textbf{\bibinfo{volume}{69}},
  \bibinfo{pages}{034007} (\bibinfo{year}{2004}).
  %, \bibinfo{note}{comments: 15  pages. Cracow Epiphany Conference January 5-7 2009}, \eprint{0905.2492},
  %\urlprefix\url{http://arxiv.org/abs/0905.2492}.


\bibitem[{\citenamefont{Gustafson}(2009)}]{Gustafson}

%\refitem{article}
\bibinfo{author}{\bibfnamefont{G.}~\bibnamefont{Gustafson}},
  \bibinfo{journal}{Acta Phys. Polon. B} \textbf{\bibinfo{volume}{40}},
  \bibinfo{pages}{1981} (\bibinfo{year}{2009}).
  %, \bibinfo{note}{comments: 15  pages. Cracow Epiphany Conference January 5-7 2009}, \eprint{0905.2492},
  %\urlprefix\url{http://arxiv.org/abs/0905.2492}.

\bibitem[{\citenamefont{Flensburg \emph{et~al.}}(2009)\citenamefont{Flensburg,
  Gustafson, and Lonnblad}}]{Lonnbland1}

%\refitem{article}
\bibinfo{author}{\bibfnamefont{C.}~\bibnamefont{Flensburg}},
  \bibinfo{author}{\bibfnamefont{G.}~\bibnamefont{Gustafson}},
  \bibnamefont{and} \bibinfo{author}{\bibfnamefont{L.}~\bibnamefont{Lonnblad}},
  \bibinfo{journal}{Eur. Phys. J. (C)} \textbf{\bibinfo{volume}{60}},
  \bibinfo{pages}{233} (\bibinfo{year}{2009}).
  %\urlprefix\url{http://arxiv.org/abs/0807.0325v1}.

\bibitem[{\citenamefont{Braun and Vechernin}(2004)}]{Braun-Vechernin}

%\refitem{article}
\bibinfo{author}{\bibfnamefont{M.~A.} \bibnamefont{Braun}} \bibnamefont{and}
  \bibinfo{author}{\bibfnamefont{V.~V.} \bibnamefont{Vechernin}},
  \bibinfo{journal}{Theor. Math. Phys.}
  \textbf{\bibinfo{volume}{139}}, \bibinfo{pages}{766} (\bibinfo{year}{2004}).
  %\urlprefix\url{http://dx.doi.org/10.1023/B:TAMP.0000029700.33219.5b}.

\bibitem[{\citenamefont{{В. В. Вечернин, Р.
  С. Колеватов }}(2004{\natexlab{a}})}]{diskr1}

%\refitem{article}
\bibinfo{author}{\bibnamefont{{В. В. Вечернин, Р. С. Колеватов}}},
  \bibinfo{journal}{Вест. СПбГУ. Сер. 4. Физика. Химия. Вып. 2, 12}
  (\bibinfo{year}{2004}{\natexlab{a}}), arXiv: hep-ph/0304295.

\bibitem[{\citenamefont{{В. В. Вечернин, Р.
  С. Колеватов}}(2004{\natexlab{b}})}]{diskr2}

%\refitem{article}
\bibinfo{author}{\bibnamefont{{В.~В.~Вечернин, Р.~С.~Колеватов}}},
  \bibinfo{journal}{Вест. СПбГУ. Сер. 4. Физика. Химия. Вып. 4, 11}
  (\bibinfo{year}{2004}{\natexlab{b}}), arXiv: hep-ph/0305136.

\bibitem[{\citenamefont{{M.~A.~Braun, R.~S.~Kolevatov, C.~Pajares, V.~V.~Vechernin}}(2004)}]{diskr3}

%\refitem{article}
\bibinfo{author}{\bibnamefont{{M.~A.~Braun, R.~S.~Kolevatov, C.~Pajares, V.~V.~Vechernin}}}, \bibinfo{journal}{Eur. Phys. J. C}
  \textbf{\bibinfo{volume}{32}}, \bibinfo{pages}{535} (\bibinfo{year}{2004}).

\bibitem[{\citenamefont{{P.A. Bolokhov, M.A. Braun, G.A. Feofilov, V.P.
  Kondratiev, V.V. Vechernin}}(2002)}]{sigmafit}

%\refitem{article}
\bibinfo{author}{\bibnamefont{{P.~A.~Bolokhov, M.~A.~Braun, G.~A.~Feofilov, V.~P.~Kondratiev,
and V.~V.~Vechernin}}}, \bibinfo{journal}{ALICE Internal Note/PHY,}
  {\bibinfo{volume}{ALICE-INT-2002-20 1.0}} (\bibinfo{year}{2002}).

\bibitem[{\citenamefont{Khachatryan
  \emph{et~al.}}(2010)\citenamefont{Khachatryan, Sirunyan, Tumasyan
  \emph{et~al.}}}]{CMSfit}

%\refitem{article}
\bibinfo{author}{\bibfnamefont{V.}~\bibnamefont{Khachatryan}},
  \bibinfo{author}{\bibfnamefont{A.~M.} \bibnamefont{Sirunyan}},
  \bibinfo{author}{\bibfnamefont{A.}~\bibnamefont{Tumasyan}},
  \bibnamefont{\emph{et~al.}} (\bibinfo{collaboration}{CMS Collab.}),
  \bibinfo{journal}{Phys. Rev. Lett.} \textbf{\bibinfo{volume}{105}},
  \bibinfo{pages}{022002} (\bibinfo{year}{2010}).
  %\urlprefix\url{http://link.aps.org/doi/10.1103/PhysRevLett.105.022002}.

\bibitem[{\citenamefont{Aamodt \emph{et~al.}}(2010{\natexlab{b}})}]{alice10}

%\refitem{article}
\bibinfo{author}{\bibfnamefont{K.~Aamodt, N.~Abel, U.~Abeysekara,}}
  \bibnamefont{\emph{et~al.}} (\bibinfo{collaboration}{ALICE Collab.}),
  \bibinfo{journal}{Eur. Phys. J. C} \textbf{\bibinfo{volume}{68}},
  \bibinfo{pages}{345} (\bibinfo{year}{2010}{\natexlab{b}}).


\bibitem[{\citenamefont{{V. Kovalenko}}(2010)}]{KovSP}

%\refitem{article}
\bibinfo{author}{\bibnamefont{{V. Kovalenko}}}, \bibinfo{journal}{\textit{in Proceedings of
 International Student Conference "Science and Progress"}},  St.~Petersburg (\bibinfo{year}{2010}), p. \bibinfo{pages}{257}.


\bibitem[{\citenamefont{{C.~Alt, et al. (NA49 Collab.) and 
G.~Feoilov, R.~Kolevatov, V.~Kondratiev, P.~Naumenko, V.~Vehernin}}(2005)}]{NA49}

%\refitem{article}
\bibinfo{author}{\bibnamefont{{C.~Alt, et al. (NA49 Collab.) and 
G.~Feoilov, R.~Kolevatov, V.~Kondratiev, P.~Naumenko, V.~Vehernin}}}, \bibinfo{journal}{\textit{in Proceedings of XVII
  International Baldin Seminar on High Energy Physics Problems}, JINR, Dubna} (\bibinfo{year}{2005}),
p. \bibinfo{pages}{222}.

\bibitem[{\citenamefont{Aamodt \emph{et~al.}}(2010{\natexlab{b}})}]{Alice900}

%\refitem{article}
\bibinfo{author}{\bibfnamefont{K.~Aamodt, N.~Abel, U.~Abeysekara,}}
  \bibnamefont{\emph{et~al.}} (\bibinfo{collaboration}{ALICE Collab.}),
  \bibinfo{journal}{Phys. Lett. B} \textbf{\bibinfo{volume}{693}},
  \bibinfo{pages}{53} (\bibinfo{year}{2010}{\natexlab{b}}), \eprint{arXiv: 1007.0719 [hep-ex]}.

\end{thebibliography}
%
% Список литературы
%
%
\newpage
%
% Для статей на русском языке далее следуют
% на английском языке название статьи, список авторов и краткая аннотация.
%
\selectlanguage{english}
\begin{center}
\large \bfseries \MakeTextUppercase{%
Modelling of exclusive parton distributions
and long-range rapidity correlations
for \textit{pp}-collisions at the LHC energy.
}
\end{center}
\begin{center}
\bfseries V.~N.~Kovalenko
\end{center}
\begin{center}
\begin{minipage}{\textwidth - 2cm}
\small
Soft pp interactions are considered in the framework of the phenomenological model with color strings formation. Under the assumption, that the elementary collision is realized as interaction of two color dipoles, the total inelastic cross section and the multiplicity of the charged particles are estimated and used for the parameters fixing. The special attention is given to the modelling of exclusive parton distributions taking into account the energy conservation and fixing the mass center that is necessary for the description of correlations, corresponding Monte Carlo procedures are developed for an arbitrary number of partons in a hadron. The string fusion algorithm in a cross-section plane is developed taking into account the finite rapidity width of strings. In this framework the influence of string fusion effects on long-range correlations of charged particles are estimated. The proposed algorithms are planned to be used for studying of heavy ion collisions.
\end{minipage}
\end{center}
\selectlanguage{russian}
\newpage
%Рисунок в статью можно включить при помощи окружения figure:

\begin{figure}[t!]
\setcaptionmargin{5mm}

%\onelinecaptionsfalse % если подпись к рисунку многострочная
\end{figure}

%список подписей к рисункам.
\newpage

\ \ 

\newpage
\setcounter{figure}{0}

\begin{figure}[H]
\setcaptionmargin{5mm}
{

\captionstyle{normal} \caption{\label{pic1lab} Множественность и средний поперечный импульс заряженных частиц в \textit{pp}-столкновениях. Линиями показаны фиты экспериментальных данных из работ \cite{alice10,CMSfit}.}
}

\end{figure}

\begin{figure}[H]
\setcaptionmargin{5mm}

%\onelinecaptionsfalse % если подпись к рисунку многострочная
\onelinecaptionstrue  % если подпись к рисунку однострочная
{
\captionstyle{normal} \caption{\label{pic2lab}
$nn$-корреляционная функция в \textit{pp}-столкновениях при энергии $E=900$~ГэВ, вычисленная
в данной работе для быстротных окон $(-0.8;0)$ и $(0.0;0.8)$.
Крестиками показана $nn$-корреляция без учета эффекта слияния струн.
}

}

\end{figure}

\begin{figure}[H]
\setcaptionmargin{5mm}

%\onelinecaptionsfalse % если подпись к рисунку многострочная
\onelinecaptionstrue  % если подпись к рисунку однострочная
{

\captionstyle{normal} \caption{\label{pic3lab}
Корреляционная функция дальних $p_tn$-корреляций в $pp$-столкновениях, вычисленная в данной работе (слева), для быстротных окон $(-0.8;0)$ и $(0.0;0.8)$;
экспериментальные данные (справа) при энергии $E=900$~ГэВ \cite{Alice900}.
Крестиками показана $p_tn$-корреляция без учета эффекта слияния струн.
}
}

\end{figure}

\end{document}